\documentclass[prc,twocolumn,showpacs,amsmath,amssymb]{revtex4}

\usepackage{times}
\usepackage{graphicx}
\usepackage{dcolumn}
\usepackage{bm}
\usepackage{amstext}

\begin{document}

\title{Relativistic mean field study of the properties of Z=117 nucleus and the decay chains of $^{293,294}$117 isotopes 
}
\author{M. Bhuyan$^{1,2}$, S. K. Patra$^{1}$ and Raj K. Gupta$^3$}

\affiliation{$^1$ Institute of Physics, Sachivalaya Marg, Bhubaneswar-751 005, India. \\ 
$^2$ School of Physics, Sambalpur University, Jyotiviha, Sambalpur-768 019, India.\\
$^3$ Department of Physics, Panjab University, Chandigarh-160 014, India.
}

\date{\today}

\begin{abstract}

We have calculated the binding energy, root-mean-square radius and quadrupole deformation parameter for the recently 
synthesized superheavy element Z=117, using the axially deformed relativistic mean field (RMF) model. The calculation 
is extended to various isotopes of Z=117 element, strarting from A=286 till A=310. We predict almost spherical structures 
in the ground state for almost all the isotopes. A shape transition appears at about A=292 from prolate to a oblate shape 
structures of Z=117 nucleus in our mean field approach. The most stable isotope (largest binding energy per nucleon) 
is found to be the $^{288}$117 nucleus. Also, the Q-value of $\alpha$-decay $Q_\alpha$ and the half-lives $T_{\alpha}$ 
are calculated for the $\alpha$-decay chains of $^{293}$117 and $^{294}$117, supporting the magic numbers at N=172 and/ 
or 184.

\end{abstract}

\pacs{21.10.Dr, 21.10.Ft, 21.10.Gv, 21.10.Tg}

\maketitle

\section{Introduction}

Nuclei can survive beyond the macroscopic limit, far into the transuranium region, where the necessary balance between the 
nuclear and Coulomb force is achieved only through shell stabilisation. Superheavy element (SHE) are hypothesised to exist 
in this region. The next double shell closer, beyond $^{208}$Pb, predicted at Z=114, N=184, may have suprisingly long 
half-life, even of the order of a million year \cite{myers65,sobi66,meldner67,mosel69,buro05}. 

Experimentally, till to-date, elements upto Z=118 have been synthesised by heavy ion reactions \cite{hof00,oga07}, with 
half-lives ranging from a few minutes to about a milli-second. The more microscopic theoretical calculations have predicted 
the next region of stability, beyond Z=82, N=126, as Z=120, N=172 or 184 \cite{rutz97,gupta97,patra1} and Z=124 or 126, 
N=184 \cite{cwiok96,kruppa00}. However, the recent experimental possibility of Z=122 from natural 
$^{211,213,217,218}$Th-isotopes, associated with long lived superdiformed (SD) and/ or hyperdeformed (HD) isomeric states 
\cite{marinov07,marinov09,bhu09}, by 16 to 22 orders of magnitude longer than their corresponding ground-state, and more 
recently the synthesis of Z=117 at Flerov Laboratory \cite{oga10} from the reaction 
$^{48}_{20}$Ca+$^{249}_{97}$Bk$\rightarrow ^{297}$117 (Z=117, A=297), which decay simultaneously via three and four 
neutrons into two differnet isotopes $^{293}$117 and $^{294}$117, motivates us to focus on their properties, using a 
microscopic theoritical model with better predictive power. Such estimations of structure properties of nuclei in the
superheavy mass region is a challenging area in nuclear physics and a fruitful path towards the understanding of 
``Island of stability" beyond the spherically doubly-magic nucleus $^{208}$Pb.

The paper is organised as follows. Section II gives a brief description of the relativistic mean field formalism. The 
pairing effects for open shell nuclei, included in our calculations, are the same as discussed in \cite{bhu09}. The results 
of our calculation are presented in Section III, and Section IV includes the $\alpha$-decay modes of $^{293}$117 and 
$^{294}$117 isotopes. A summary of our results, together with the concluding remarks, are given in the last Section V. 

\section{The relativistic mean-field (RMF) method}

The relativistic Lagrangian density for a nucleon-meson many-body system \cite{sero86,ring90},
\begin{eqnarray}
{\cal L}&=&\overline{\psi_{i}}\{i\gamma^{\mu}
\partial_{\mu}-M\}\psi_{i}
+{\frac12}\partial^{\mu}\sigma\partial_{\mu}\sigma
-{\frac12}m_{\sigma}^{2}\sigma^{2}\nonumber\\
&& -{\frac13}g_{2}\sigma^{3} -{\frac14}g_{3}\sigma^{4}
-g_{s}\overline{\psi_{i}}\psi_{i}\sigma-{\frac14}\Omega^{\mu\nu}
\Omega_{\mu\nu}\nonumber\\
&&+{\frac12}m_{w}^{2}V^{\mu}V_{\mu}
+{\frac14}c_{3}(V_{\mu}V^{\mu})^{2} -g_{w}\overline\psi_{i}
\gamma^{\mu}\psi_{i}
V_{\mu}\nonumber\\
&&-{\frac14}\vec{B}^{\mu\nu}.\vec{B}_{\mu\nu}+{\frac12}m_{\rho}^{2}{\vec
R^{\mu}} .{\vec{R}_{\mu}}
-g_{\rho}\overline\psi_{i}\gamma^{\mu}\vec{\tau}\psi_{i}.\vec
{R^{\mu}}\nonumber\\
&&-{\frac14}F^{\mu\nu}F_{\mu\nu}-e\overline\psi_{i}
\gamma^{\mu}\frac{\left(1-\tau_{3i}\right)}{2}\psi_{i}A_{\mu} .
\end{eqnarray}
All the quantities have their usual well known meanings. From the above Lagrangian we obtain the field equations for the 
nucleons and mesons. These equations are solved by expanding the upper and lower components of the Dirac spinors and the 
boson fields in an axially deformed harmonic oscillator basis, with an initial deformation $\beta_{0}$. The set of coupled 
equations is solved numerically by a self-consistent iteration method. The centre-of-mass motion energy correction is 
estimated by the usual harmonic oscillator formula $E_{c.m.}=\frac{3}{4}(41A^{-1/3})$. The quadrupole deformation parameter 
$\beta_2$ is evaluated from the resulting proton and neutron quadrupole moments, as 
$Q=Q_n+Q_p=\sqrt{\frac{16\pi}5} (\frac3{4\pi} AR^2\beta_2)$. The root mean square (rms) matter radius is defined as 
$\langle r_m^2\rangle={1\over{A}}\int\rho(r_{\perp},z) r^2d\tau$, where $A$ is the mass number, and $\rho(r_{\perp},z)$ is 
the deformed density. The total binding energy and other observables are also obtained by using the standard relations, 
given in \cite{ring90}. We use the well known NL3 parameter set \cite{lala97}. This set reproduces the properties of not 
only the stable nuclei but also well predicts for those far from the $\beta$-stability valley. As outputs, we obtain 
different potentials, densities, single-particle energy levels, radii, deformations and the binding energies. For a given 
nucleus, the maximum binding energy corresponds to the ground state and other solutions are obtained as various excited 
intrinsic states. 

The constant gap, BCS-pairing approach is reasonably valid for nuclei in the valley of $\beta$-stability line. However, 
this method breaks down when the coupling of the continuum becomes important. In the present study, we deal with nuclei on 
or near the valley of stability line since the superheavy elements, though very exotic in nature, lie on the 
$\beta$-stability line. In order to take care of the pairing effects in the present study, we use the constant gap for 
proton and neutron, as given in \cite{madland81}, which are valid for nuclei both on or away from the stability line 
(for more details, see, e.g., Ref. \cite{bhu09}, where $E_{pair}$, the pairing energy, is also defined).    

\section{Results and Discussion}

In many of our previous works and of other authors \cite{patra1,ring90,lala97,patra2,patra3,patra4}, the ground state 
properties, like the binding energies (BE), pairing energies $E_{pair}$, quadrupole deformation parameters $\beta_2$, 
charge radii ($r_{ch}$), and other bulk properties, are evaluated by using the above stated relativistic Langragian for 
different forces. From these predictions, it is found that, generally speaking, most of the recent parameter sets reproduce 
well the ground state properties, not only of stable normal nuclei but also of exotic nuclei far away from the valley of 
$\beta$-stability. This means to say that if one uses a reasonably well accepted parameter set, the predictions of the 
model will remain nearly force independent. In this paper we have used the successful NL3 parameter sets for our 
calculation. 

\begin{figure}[ht]
\vspace{0.75cm}
\begin{center}
\includegraphics[width=1.0\columnwidth]{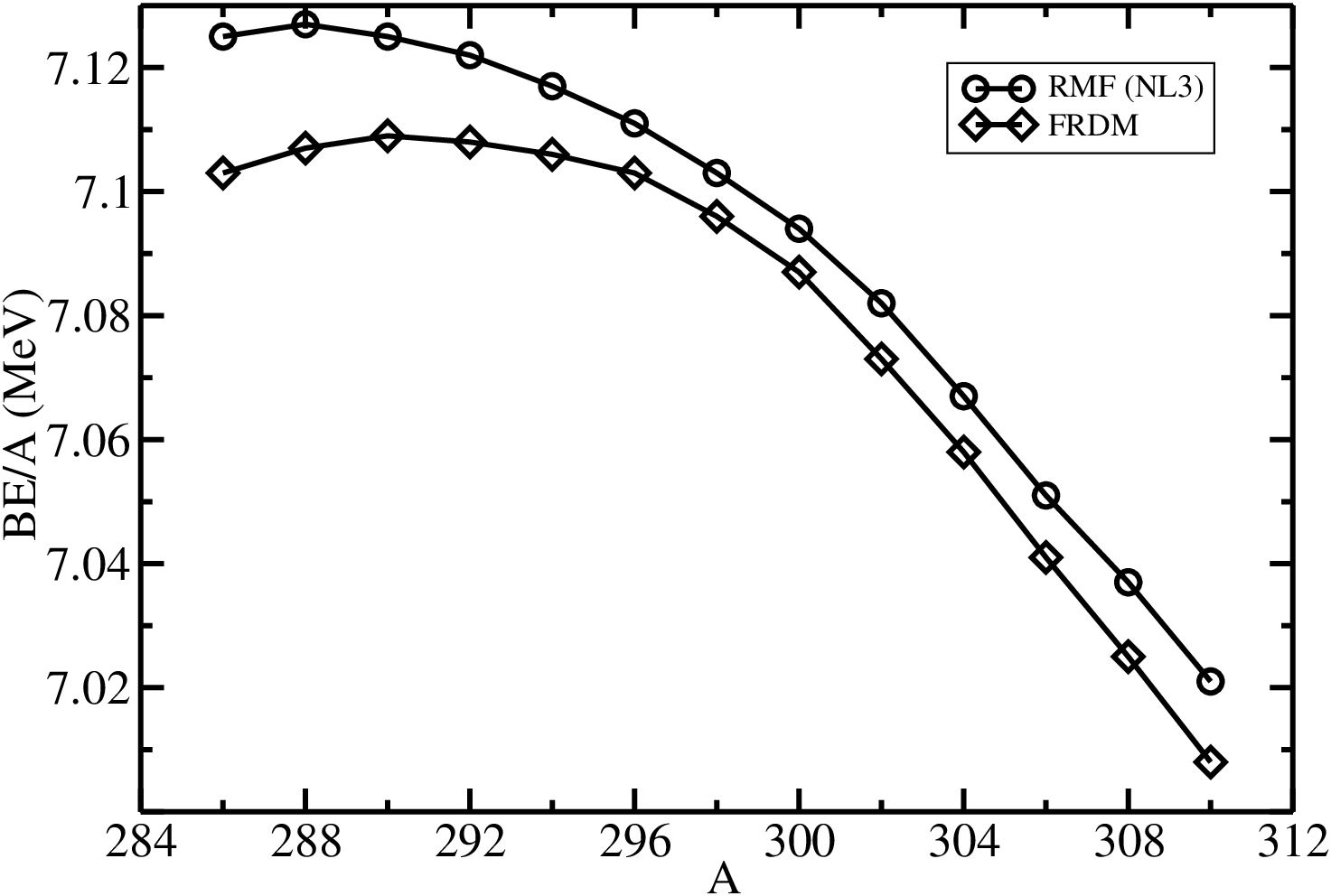}
\caption{The binding energy per particle BE/A for the $^{284-310}$117 isotopes, obtained in RMF(NL3) formalism and compared 
with the FRDM results \cite{moll97}, wherever available.
}
\end{center}
\label{Fig. 1}
\end{figure}

\begin{figure}[ht]
\begin{center}
\includegraphics[width=1.0\columnwidth]{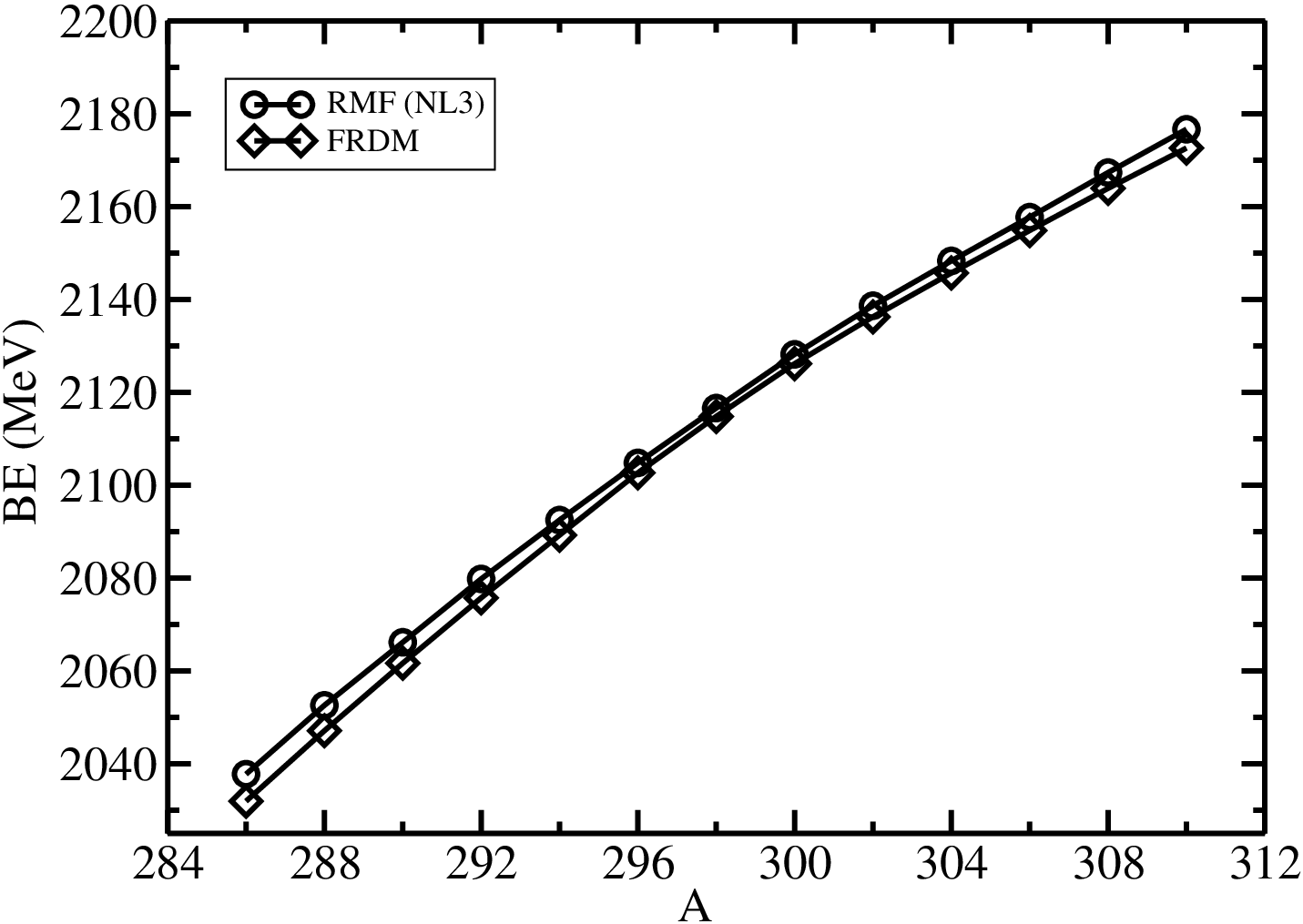}
\caption{The total binding energy BE for $^{284-310}$117 nuclei in RMF(NL3) compared with the FRDM results \cite{moll97}.
}
\end{center}
\label{Fig. 2}
\end{figure}

\begin{table*}
\caption{The RMF(NL3) results for binding energy BE, two-neutron separation energy $S_{2n}$, pairing energy $E_{pair}$, 
the binding energy difference $\triangle E$ between the ground- and first-exited state solutions, and the quadrupole 
deformation parameter $\beta_{2}$, compared with the corresponding Finite Range Droplet Model (FRDM) results \cite{moll97}. 
The energy is in MeV.
}
\begin{tabular}{|c|c|c|c|c|c|c|c|c|}
\hline
&\multicolumn{5}{c|}{RMF(NL3) Result}&\multicolumn{3}{c|}{FRDM Result}\\
\hline
Nucleus& BE  & $S_{2n}$ & $E_{pair}$ & $\Delta E$ & $\beta_{2}$ &  BE  & $S_{2n}$ & $\beta_{2}$ \\
\hline
288 & 2052.586 & 14.836 & 14.698 & 0.333 & 0.018 & 2047.09 & 15.16 & 0.080 \\   
290 & 2066.138 & 13.552 & 14.274 & 0.360 & 0.017 & 2061.65 & 14.56 & 0.080 \\ 
292 & 2079.802 & 13.664 & 14.109 & 0.096 & -0.017 & 2075.72 & 14.07 & 0.072 \\
294 & 2092.468 & 12.775 & 13.653 & 0.031 & 0.041 & 2089.22 & 13.50 & -0.087 \\
296 & 2104.803 & 12.335 & 13.583 & 0.104 & 0.028 & 2102.66 & 13.45 & -0.035 \\
298 & 2116.598 & 11.691 & 13.274 & 0.389 & 0.015 & 2114.79 & 12.13 & -0.008 \\
300 & 2128.174 & 11.576 & 12.841 & 0.970 & 0.005 & 2126.14 & 11.34 & 0.000  \\
302 & 2138.662 & 10.488 & 12.623 & 0.596 & 0.004 & 2136.25 & 10.11 & 0.000  \\
304 & 2148.296 & 9.634  & 12.695 & 0.012 & 0.002 & 2145.71 & 9.46  & 0.000  \\
306 & 2157.726 & 9.430  & 12.348 & 0.004 & 0.030 & 2154.84 & 9.13  & 0.000  \\
308 & 2167.327 & 9.601  & 11.912 & 0.304 & 0.047 & 2163.93 & 9.09  & 0.001  \\
310 & 2176.656 & 9.329  & 11.538 & 0.512 & 0.051 & 2172.61 & 8.68  & 0.000  \\
\hline
\end{tabular}
\label{Table 1}
\end{table*}

\begin{figure}[ht]
\vspace{0.2cm}
 \begin{center}
\includegraphics[width=1.0\columnwidth]{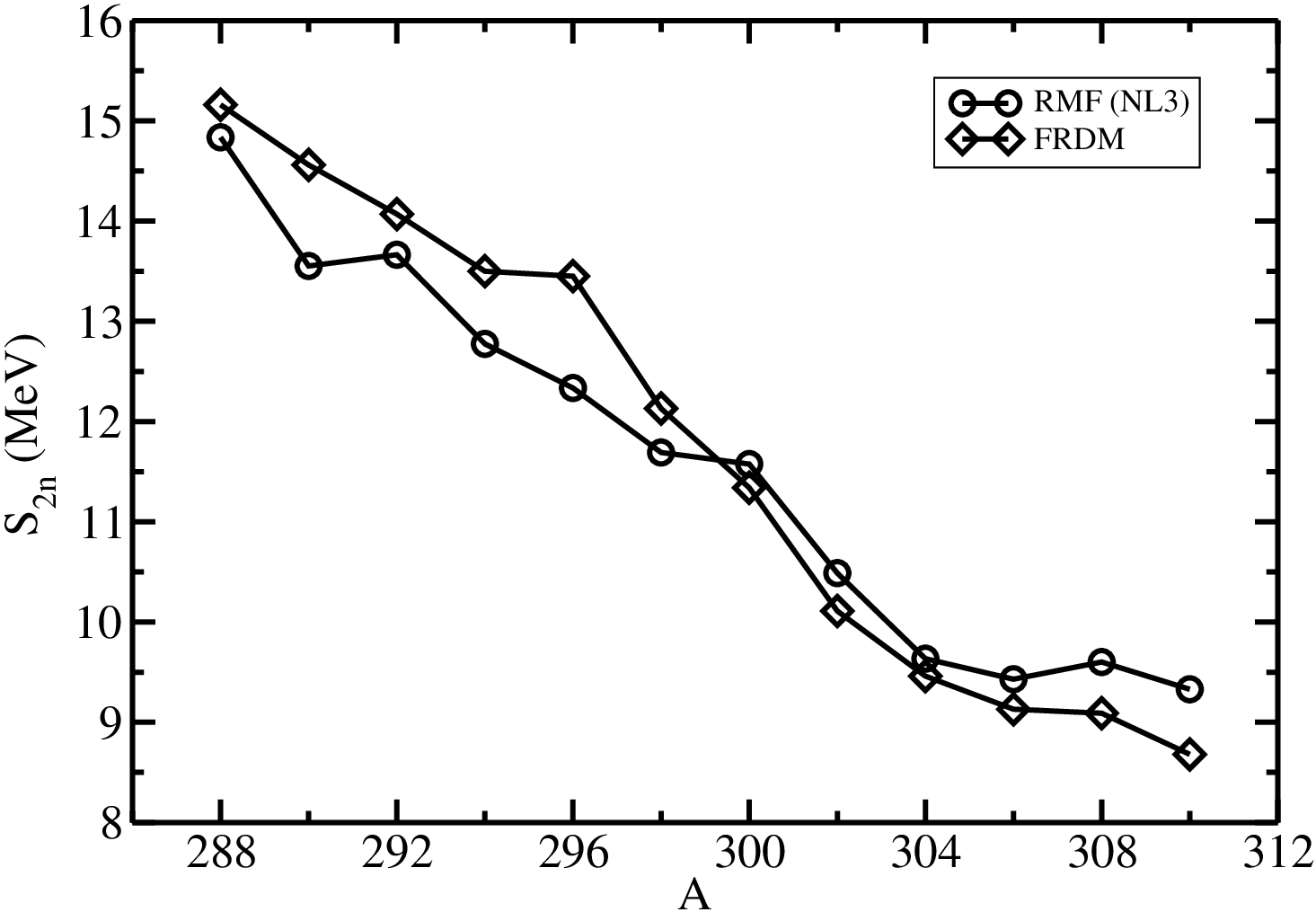}
\caption{The two-neutron separation energy ${S_2n}$ for $^{286-310}$117 
nuclei, obtained from RMF(NL3) formalisms, and compared with the FRDM 
results \cite{moll97}, wherever available.}                              
\end{center}
\label{Fig. 3}
\end{figure}

\subsection{Binding energy and two-neutron separation energy}

The calculated binding energy per nucleon BE/A and binding energy BE, obtained from the RMF(NL3) formalism, are comapred, 
respectively, in Figs. 1 and 2 and in Table I, with the Finite Range Droplet Model (FRDM) results \cite{moll97}. We notice 
that the BE/A obtained in the RMF(NL3) model over-estimate the FRDM result. In general, the BE/A value starts increasing 
with the increase of mass number A, reaching a peak value at A=288 for RMF(NL3) and at A=290 for the FRDM formalism. This 
means to say that $^{288}$117 is the most stable isotope from the RMF(NL3) result and $^{290}$117 from the FRDM 
predictions. Interestingly, $^{288}$117 (with N=171) and  $^{290}$117 (with N=173) are both closer to the predicted closed 
shell at N=172 than at N=184. Note that the isotopes $^{300, 302}$117, next to the magic number N=184, are also included 
in this study. For the total binding energy BE of the isotopic chain in Table I and Fig. 2, we notice that the microscopic 
RMF binding energies agree well with the macro-microscopic FRDM calculations, their differences decreasing gradually 
towards the higher mass region (around A=298), and then beyond this mass number the two curves again showing a similar 
behavior. Note that $^{298}$117 (with N=181) is in this case closer to N=184.

The two-neutron separation energy $S_{2n}$(N,Z)= BE(N,Z)-BE(N-2,Z) is also listed in Table I. From the table, we find that 
the microscopic RMF $S_{2n}$ values also agree well with the macro-microscopic FRDM calculations. This comparison of 
$S_{2n}$ for RMF with the FRDM result are further shown in Fig. 3, which clearly shows that the two $S_{2n}$ values 
coincide remarkably well, except at masses A=290 and 296. Apparently, the $S_{2n}$ decrease gradually with increase of 
neutron number, except for the noticeable kinks at A=290 (with N=173) and A=300 (with N=183) in RMF, and at A=296 (with 
N=179) in FRDM.  Interestingly, these neutron numbers for RMF(NL3) are close to the earlier predicted 
\cite{rutz97,gupta97,patra1} N=172 or 184 magic numbers.
  
\subsection{Shape co-existence}

We have also calculated, for the whole Z=117 isotopic chain, the solutions in both prolate and oblate deformed 
configurations. In many cases, we find low lying excited states. As a measure of the energy difference between the ground 
band-head and the first excited state, we have plotted in Fig. 4 (a) the binding energy difference $\triangle E$ between 
the two solutions, noting that the maximum binding energy solution refers to the ground state (g.s.) and all other 
solutions to the intrinsic excited state (e.s.). From Fig. 4 (a), we notice that in RMF calculations, the energy difference 
$\triangle E$ is small for the whole region of the considered isotopic series. This small difference in the binding energy 
for neutron-deficient isotopes is an indication of the presence of shape co-existence. In other words, the two solutions 
in these nuclei are almost degenerate for a small difference of output in energy. For example, in $^{290}$117, the two 
solutions for $\beta_2$=0.017 and 0.123 are completely degenerate with the binding energies 2066.138 and 2065.778 MeV. This 
later result means to suggest that the ground state can be changed to the excited state, and vice-versa, by a small change 
in the input data, like the pairing strength, etc., in the calculations. In any case, such a phenomenon is known to exist 
in many other regions of the periodic table \cite{patra94}.

\begin{figure}[ht]
\vspace{0.65cm}
\begin{center}
\includegraphics[width=1.0\columnwidth]{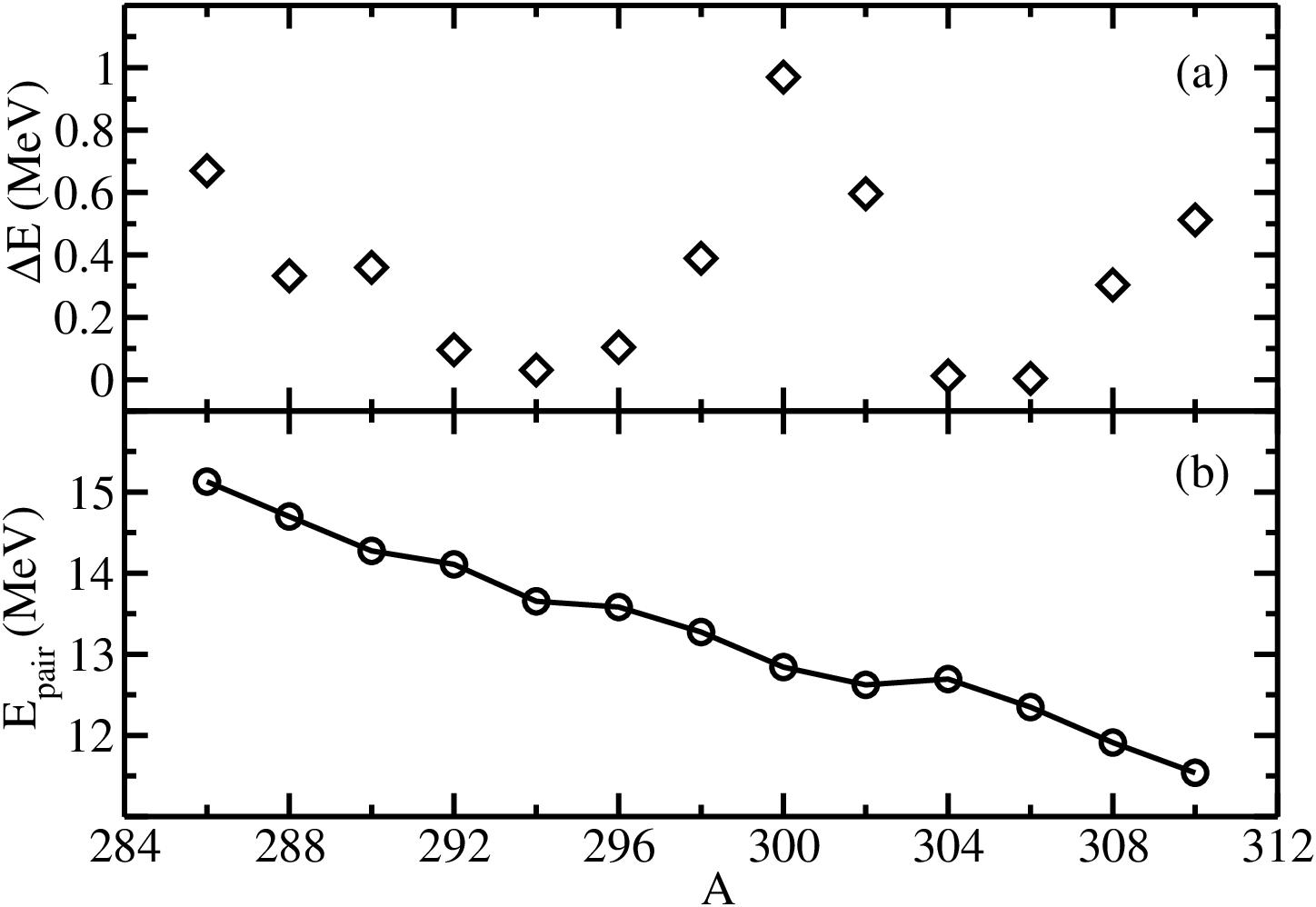}
\caption{(a) The energy difference between the ground-state and the first-excited state $\triangle E$, and (b) the pairing 
energy $E_{pair}$, for the relativistic RMF(NL3) calculation of Z=117 isotopic chain.
}
\end{center}
\label{Fig. 4}
\end{figure}

Pairing is important for open shell nuclei whose value, for a given nucleus, depends only marginally on quadrupole 
deformation $\beta_2$. This means that for differing $\beta_2$-values in a nucleus, the pairing energy $E_{pair}$ changes 
only marginally (by $\sim$5-6$\%$). On the other hand, even if the $\beta_2$ values for two nuclei are same, the 
$E_{pair}$'s could be different from one another, depending on the filling of the nucleons. This result is illustrated in 
Fig. 4 (b) for the RMF(NL3) calculation, where $E_{pair}$ for both the g.s. and first-excited state (e.s.), refering to 
different $\beta_2$-values, are plotted for the full isotopic chain. It is clear from Fig. 4(b) that $E_{pair}$ decreases 
with increase in mass number A, i.e., even if the $\beta_2$ values for two nuclei are the same, the $E_{pair}$'s are 
different from one another. This change of $E_{pair}$ is $\sim$15$\%$ in going from, say, A=286 to 310. 

\begin{figure}[ht]
\vspace{0.40cm}
\begin{center}
\includegraphics[width=1.0\columnwidth]{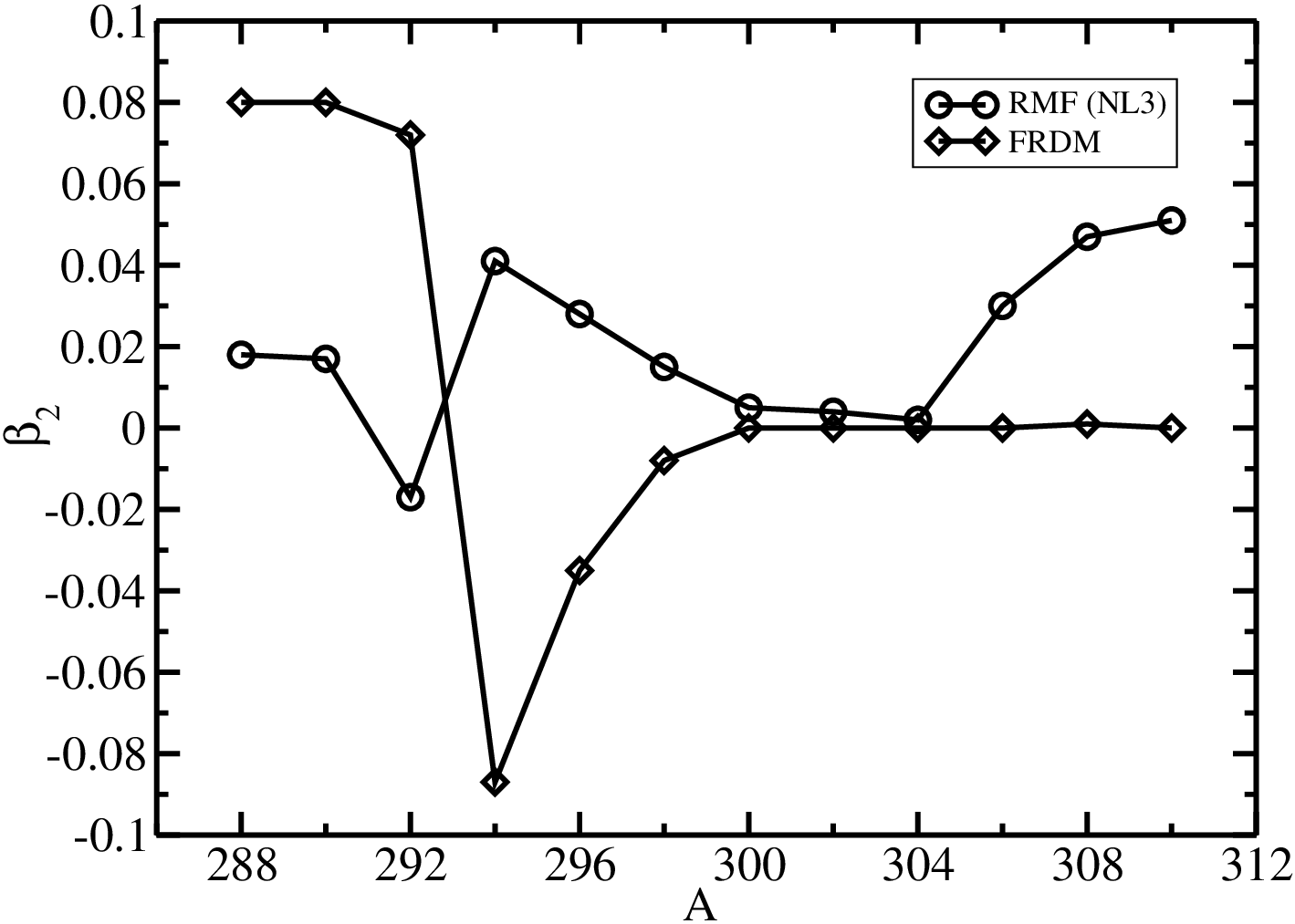}
\caption{Quadrupole deformation parameter obtained from relativistic mean field formalism RMF(NL3), Compared with the FRDM 
results \cite{moll97}, wherever available.
}
\end{center}
\label{Fig. 5}
\end{figure}

\subsection{Quadrupole deformation parameter}

The quadrupole deformation parameter $\beta_2$, for both the ground and first excited states, are also determined within 
the RMF formalism. In some of the earlier RMF and Skyrme Hartree-Fock (SHF) calculations, it was shown that the quadrupole 
moment obtained from these theories reproduce the experimental data pretty well 
\cite{patra1,sero86,ring90,lala97,patra2,rei95,cha97,cha98,brown98}. The g.s. quadrupole deformation parameter 
$\beta_2$ is plotted in Fig. 5 for RMF, and compared with the FRDM results \cite{moll97}. It is clear from this figure that 
the FRDM results differ from the RMF(NL3) results for some mass regions. For example, the g.s. oblate solution appear for 
the nucleus $^{292}$117 in RMF but is a prolate solution in FRDM. A more careful inespection shows that the solutions 
for the whole isotopic chain are prolate, except at A=292 for RMF and at A=294-298 for FRDM model. In other word, there is 
a shape change from prolate to oblate at A=292 for RMF and at A=294 for FRDM. Interestingly, most of the isotopes are 
almost spherical in their g.s. configurations. 

\begin{figure}[ht]
\vspace{0.55cm}
\begin{center}
\includegraphics[width=1.0\columnwidth]{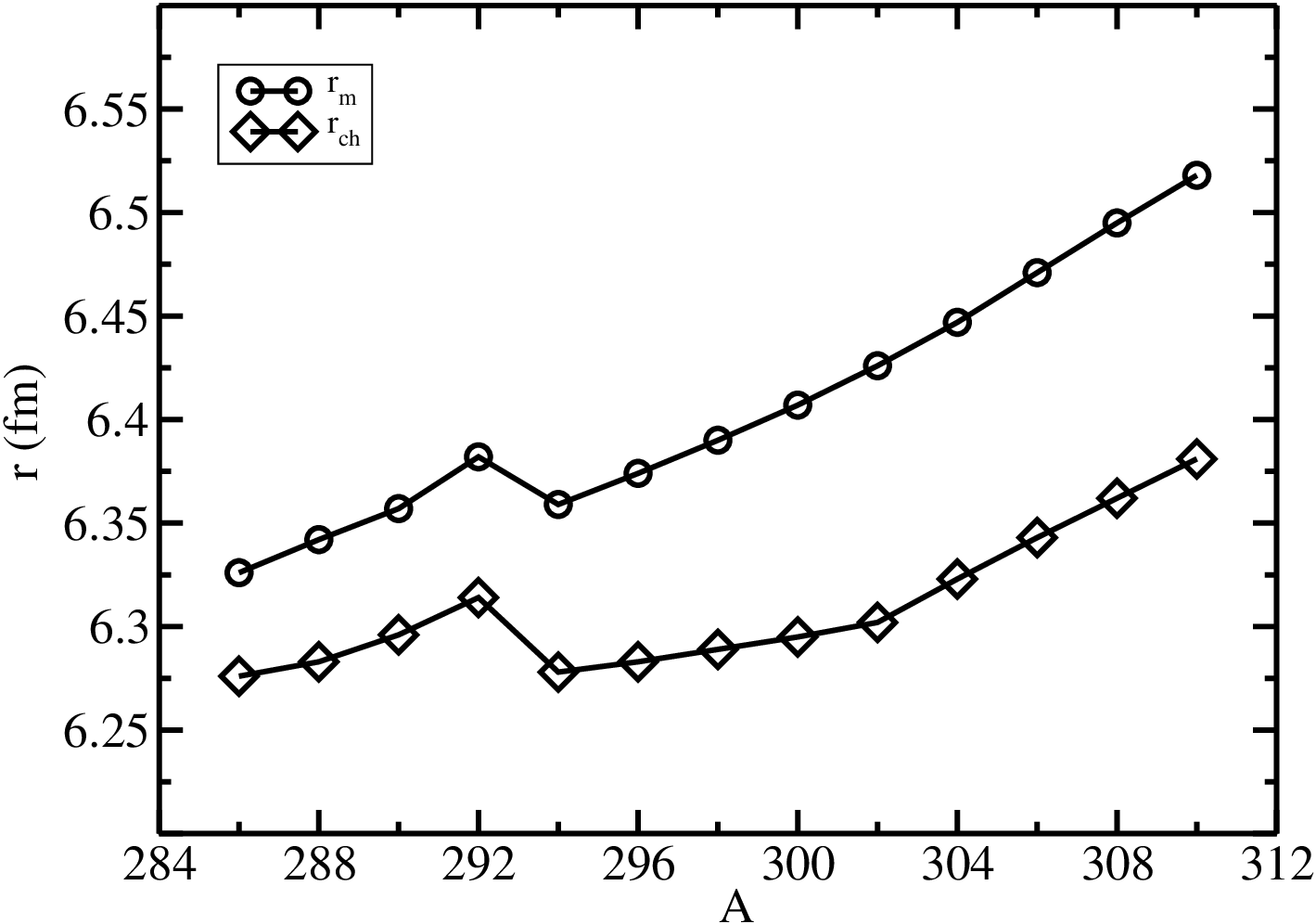}
\caption{The rms radii $r_{m}$ of  matter distribution and charge radii $r_{ch}$ for $^{286-310}$117 nuclei, using the
relativistic mean field formalism RMF(NL3).
}
\end{center}
\label{Fig. 6}
\end{figure}

\subsection{Nuclear radii}

The root-mean-square matter radius ($r_{m}$) and charge radius ($r_{ch}$) for the RMF(NL3) formalism are shown in Fig. 6. 
As expected, the matter distribution radius $r_{m}$ increases with increase of the neutron number. However, though the 
proton number Z=117 is constant for the isotopic series, the $r_{ch}$ value also increases with neutron number. A detailed 
inspection of Fig. 6 shows that, in the RMF calculations, both the radii show the monotonic increase of radii till A=310,
with a jump to a lower value at A=292 (with N=175). There is no data or other calculation available for comparisons.
 
\begin{figure}[ht]
\vspace{1.0cm}
\begin{center}
\includegraphics[width=1.0\columnwidth]{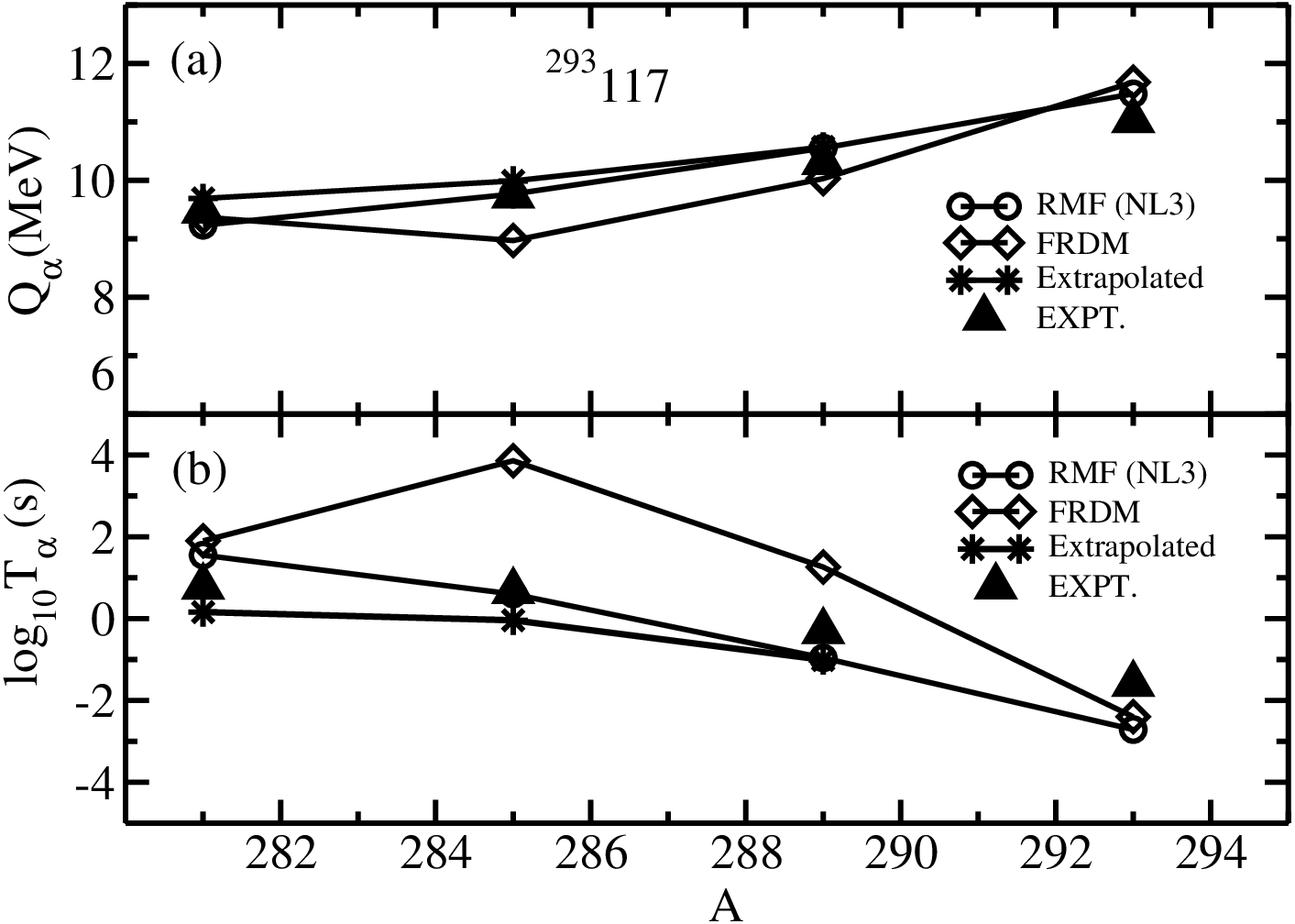}
\caption{(a) The $Q_{\alpha}$-energy and (b) the half-life time $T_{\alpha}$ for $\alpha$-decay series of $^{293}$117 
nucleus, using the relativistic mean field formalism RMF(NL3), compared with the FRDM data \cite{moll97}, the extrapolated 
experimental data \cite{ogan05,audi03} and the experimental data \cite{oga10}, whereever available.
}
\end{center}
\label{Fig. 7}
\end{figure}

\begin{figure}[ht]
\vspace{0.3cm}
\begin{center}
\includegraphics[width=1.0\columnwidth]{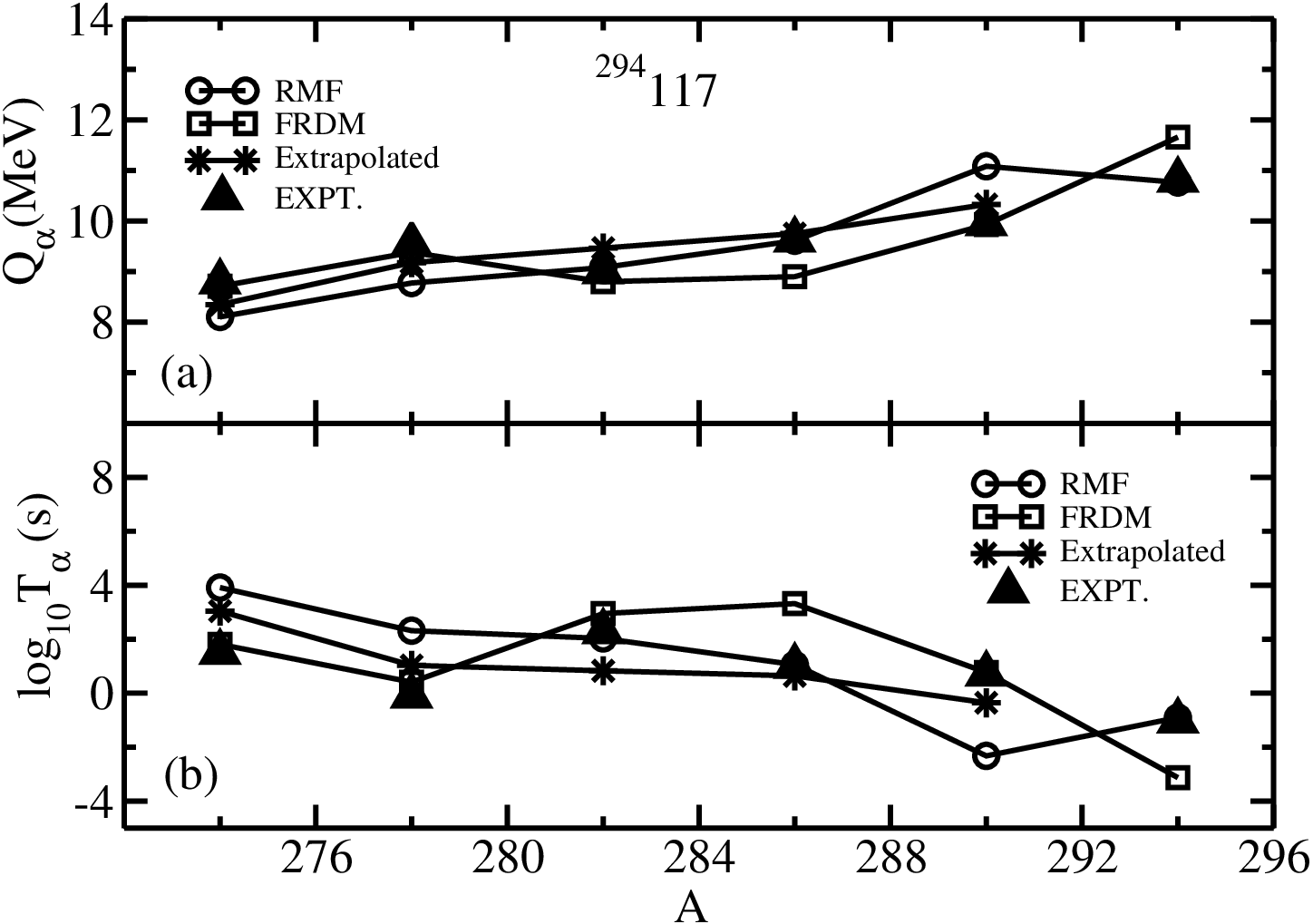}
\caption{Same as for Fig. 7, but for $^{294}$117 nucleus.
}
\end{center}
\label{Fig. 8}
\end{figure}

\begin{table*}
\caption{The $Q_{\alpha}$ energy and $T_{\alpha}$ for $\alpha$-decay series of $^{293}$117 nucleus, calculated on the 
RMF(NL3) model, and compared with the Finite Range Droplet Model (FRDM) results \cite{moll97}, the extrapolated data 
\cite{ogan05,audi03} and the experimental data \cite{oga10}, wherever available. The energy is in MeV and the half-life 
time in second.
}
\begin{tabular}{|c|c|c|c|c|c|c|c|c|c|c|c|c|}
\hline
&&\multicolumn{3}{c|}{RMF (NL3) Result} & \multicolumn{3}{c|}{FRDM Results} & \multicolumn{3}{c|}{Extrapolated result} 
& \multicolumn{2}{c|}{Experimental Results} \\
\hline
Nucleus & Z & BE & $Q_{\alpha}$ & $T_{\alpha}$ & BE & $Q_{\alpha}$ & $T_{\alpha}$& BE & $Q_{\alpha}$ & $T_{\alpha}$ 
& $Q_{\alpha}$ & $T_{\alpha}$\\
\hline
293 & 117 & 2086.602 & 11.480 & -2.71  & 2083.06 & 11.68 & -2.40 &         &       &       & 11.03 & -1.60 \\
289 & 115 & 2069.786 & 10.552 & -0.96  & 2066.45 & 10.03 & 1.26  & 2063.17 & 10.57 & -1.01 & 10.31 & -0.31 \\
285 & 113 & 2052.042 & 9.765  &  0.60  & 2048.18 & 8.97  & 3.86  & 2045.45 & 9.99  & -0.04 & 9.74  & 0.67 \\
281 & 111 & 2033.511 & 9.231  &  1.55  & 2028.85 & 9.37  & 1.90  & 2027.13 & 9.69  & 0.16  & 9.48  & 0.78 \\
\hline
\end{tabular}
\end{table*}

\begin{table*}
\caption{Same as for Tabe II, but for $^{294}$117 nucleus.
}
\begin{tabular}{|c|c|c|c|c|c|c|c|c|c|c|c|c|}
\hline
&&\multicolumn{3}{c|}{RMF (NL3) Result} & \multicolumn{3}{c|}{FRDM Results} & \multicolumn{3}{c|}{Extrapolated result} 
& \multicolumn{2}{c|}{Experimental Results} \\
\hline
Nucleus & Z & BE & $Q_{\alpha}$ & $T_{\alpha}$ & BE & $Q_{\alpha}$ & $T_{\alpha}$& BE & $Q_{\alpha}$ & $T_{\alpha}$ 
& $Q_{\alpha}$ & $T_{\alpha}$\\
\hline
294  & 117  & 2092.578  & 10.763 & -0.906  & 2089.22  & 11.66 & -3.13  &           &       &        & 10.81 &  -1.03 \\
290  & 115  & 2075.045  & 11.083 & -2.325  & 2072.59  & 9.93  & 0.77   & 2069.730  & 10.329& -0.358 & 9.95  &  0.714 \\
286  & 113  & 2057.832  & 9.612  & 1.055   & 2054.22  & 8.90  & 3.32   & 2051.764  & 9.752 & 0.639  & 9.63  &  1.002 \\
282  & 111  & 2039.148  & 9.080  & 2.027   & 2034.82  & 8.80  & 2.95   & 2033.220  & 9.464 & 0.830  & 9.00  &  2.286 \\
278  & 109  & 2019.932  & 8.775  & 2.316   & 2015.32  & 9.38  & 0.41   & 2014.388  & 9.176 & 1.028  & 9.55  &  -0.099 \\
274  & 107  & 2000.411  & 8.105  & 3.913   & 1996.40  & 8.71  & 1.81   & 1995.268  & 8.348 & 3.042  & 8.80  &  1.517 \\
\hline
\end{tabular}
\end{table*}

\section{The $Q_{\alpha}$ energy and the decay half-life $T_{\alpha}$}

The $Q_{\alpha}$ energy is obtained from the relation \cite{patra23}:
$$Q_{\alpha}(N, Z)=BE(N, Z)-BE(N-2, Z-2)-BE(2, 2).$$ 
Here, $BE(N, Z$) is the binding energy of the parent nucleus with neutron number N and proton number Z, $BE(2, 2)$ is the 
binding energy of the $\alpha$-particle ($^4$He), i.e., 28.296 MeV, and $BE(N-2, Z-2)$ is the binding energy of the 
daughter nucleus after the emission of an $\alpha$-particle.

With $Q_{\alpha}$ energy at hand, we esimate the half-life $log_{10}T_{\alpha}(s)$ by using the phenomenological formula 
of Viola and Seaborg \cite{viol01}:
$$log_{10}T_{\alpha}(s)=\frac {aZ-b}{\sqrt{Q_{\alpha}}}-(cZ+d).$$
Here, Z is the atomic number of parent nucleus, and $a$=1.66175, $b$=8.5166,
$c$=0.20228 and $d$=33.9069. 

\subsection{The $\alpha$-decay series of $^{293}$117 nucleus}

The binding energies of the parent and daughter nuclei are obtained by using the RMF formalism. From these BE, we evalute 
the $Q_{\alpha}$ energy and the half-life time $log_{10}T_{\alpha}(s)$, using the above formulae. Our predicted results 
for the decay chain of $^{293}117$ are compared in Table  II with the finite range droplet model (FRDM) calculation 
\cite{moll97}, the extrapolated \cite{ogan05,audi03} and the experimental data \cite{oga10}, wherever possible. The same
comparision is also carried out in Figs. 7 (a) and 7 (b), respectively, for  $Q_{\alpha}$ energy and the half-life time 
$log_{10}T_{\alpha}(s)$.  

From Figs. 7(a) and (b), and Table II, we notice that the calculated values for both $Q_{\alpha}$ and $T_{\alpha}(s)$ 
agree well with the known extrapolated as well as experimental data, but are over-estimated with-respect-to the FRDM 
predictions. For example, the values of $T_{\alpha}$ from RMF coincide well for the whole mass region with the available 
experimental data, and with extrapolated values for $^{281}$Rg, $^{285}$113 and $^{289}$115 nuclei. Similarly, for 
$^{281}Rg$, the FRDM predictions match both the extrapolated and experimental results. Furthermore, the possible shell 
structure effects in $Q_{\alpha}$, as well as in $T_{\alpha}(s)$, are noticeable for the daughter nucleus $^{285}$113 
(with N=172) for both the RMF predictions and experimental data. Note that N=172 refers to the predicted magic number.

\subsection{The $\alpha$-decay series of $^{294}$117 nucleus}

In this subsection, we present the $Q_{\alpha}$ and the $log_{10}T_{\alpha}(s)$ results for decay series of $^{294}$117 
nucleus, using the same procedure as in the previous subsection for $^{293}$117. The results obtained are listed in 
Table III and plotted in  Figs. 8(a) and 8(b), compared with the FRDM predictions \cite{moll97}, the extrapolated 
\cite{ogan05,audi03} and experimental data \cite{oga10}, wherever possible.  

From Fig. 8(a) and (b), and Table III, we found almost similar results as are predicted in the previous subsection for 
$^{293}117$. Thus, the RMF(NL3) results for both $Q_{\alpha}$ and $T_{\alpha}(s)$ agree well with the known extrapolated 
and experimental data, but once again over-estimate the FRDM results. For example, the $T_{\alpha}$ values for RMF 
coincide well with the experimental data for the whole isotopic chain and with the extrapolated data for $^{290}$115, 
$^{286}$113, $^{282}$Rg, $^{278}$Mt and $^{274}$Bh nuclei. Similarly, FRDM predictions for $^{278}$Mt and $^{290}$115 
match the extrapolated and for $^{274}$Bh and $^{278}$Mt with experimental results. The possible shell structure effects 
in $Q_{\alpha}$, as well as in $T_{\alpha}(s)$, are noticed for the daughter nucleus $^{286}$113 (with N=173) for RMF and 
$^{278}$Mt (with N=169) in experimental data, again coinciding with earlier predicted N=172 magic number.

\section{Summary}

Summarizing, we have calculated the binding energy, the rms charge and matter radii, and quadrupole deformation parameter 
for the isotopic chain of recently synthesized Z=117 superheavy element for both the ground- as well as intrinsic 
first-excited states, using the RMF formalism. From the calculated binding energy, we have also estimated the two-neutron 
separation energy and the energy diffence between ground- and first-excited state for studying the shape co-existence, 
for the isotopic chain. Also, we have estimated the pairing energy for the ground-state solution in the whole isotopic 
chain. We found a shape change from oblate to prolate deformation, with increase of isotopic mass number, at A=292. 
Most of the ground-state structures are with spherical solutions, in agreement with the FRDM calculations. From the binding 
energy analysis, we found that the most stable isotope in the series is $^{288}$117, which is close to predicted magic 
number at N=172. Our predicted $\alpha$-decay energy $Q_{\alpha}$ and half-life time $T_{\alpha}$ match nicely with the 
available extrapolated and experimental data. Some shell structure is also observed in the calculated quantities at N=172 
and/ or 184 from RMF calculations of the various isotopes of Z=117 nucleus.

\section*{Acknowledgments}

We thanks to Mr. B. B. Pani and Mr. B. K. Sahu for discussion. This work is supported in part by the UGC-DAE Consortium for 
Scientific Research, Kolkata Center, Kolkata, India (Project No. UGC-DAE CRS/KC/CRS/2009/NP06/1354).

\end{document}